\documentclass[12pt]{iopart}

\pdfoutput=1
 
\usepackage{latexsym}
\usepackage{graphics}
\usepackage{graphicx}
\usepackage{amssymb}
\usepackage{bm}
 
\begin{document}

\title{AC-Stark shift and photoionization of Rydberg atoms in an optical dipole trap}

\author{F. Markert$^2$, P. W{\"u}rtz$^2$, A. Koglbauer$^1$, T. Gericke$^2$, A. Vogler$^2$, and H. Ott$^2$}

\address{$^1$Institut f{\"u}r Physik, Johannes Gutenberg-Universit{\"a}t, 55099 Mainz, Germany \\ $^2$Research Center OPTIMAS, Technische Universit{\"a}t Kaiserslautern, 67663 Kaiserslautern, Germany}
%\mail{e-mail: ott@uni-mainz.de}

\ead{ott@physik.uni-kl.de}

\begin{abstract}
We have measured the AC-Stark shift of the $14D_{5/2}$ Rydberg state of rubidium 87 in an optical dipole trap formed by a focussed CO$_2$-laser. We find good quantitative agreement with the model of a free electron experiencing a ponderomotive potential in the light field. In order to reproduce the observed spectra we take into account the broadening of the Rydberg state due to photoionization. The extracted cross-section is compatible with previous measurements on neighboring Rydberg states. 
\end{abstract}

\maketitle

\section{Introduction}
The strong dipole-dipole interaction between Rydberg atoms has inspired many proposals for their use in quantum simulation and quantum information processing \cite{Jaksch2000,Lukin2002,Saffman2010}. Rapid experimental progress is accompanying this development \cite{Heidemann2007,Reetz-Lamour2008,Urban2009,Gaetan2009,Tauschinsky2010}. So far, the excitation has always been resonant which results in dynamics of the internal states that are much faster than the external motion of the atoms. This regime is referred to as the frozen Rydberg gas. A trapping potential is only needed in order to prepare and detect the atoms. During the internal dynamics, the trapping fields are often switched off (dipole trap) or do not play a major role (magnetic trap). 

Recently, increasing interest focuses on dressed Rydberg states \cite{Pupillo2010,Henkel2010,Honer2010,Johnson2010}. The coupling to the Rydberg state is off-resonant such that the ground state only acquires a small admixture of the Rydberg state. This helps reducing the energy scale of the dipole-dipole interaction to the same order of magnitude as the interaction between two ground state atoms and the typical kinetic energy of atoms in an ultracold sample. As the Rydberg-dressed states allow for experiments on a much longer time-scale, the frozen Rydberg gas assumption does no longer hold and effects arising from the trapping potential will be much more pronounced. It is therefore necessary to understand the role of the trapping potential and its influence on the energy levels and the lifetime of the Rydberg states.

The physical properties of the ground state of an atom can differ substantially from those of its Rydberg states. The magnetic moment does not depend on the principal quantum number $n$ but only on the angular momentum and the spin of the electrons. Typical Rydberg states that are used in current experiments ({\it ns}, {\it np}, {\it nd} states) therefore have a magnetic moment which is comparable to that of the ground state and no drastic effect is expected. Optical trapping fields instead couple to the dynamic polarizability of the atom and this can lead to significantly different light shifts for the Rydberg state compared to the ground state: the shift can be larger, smaller or can even change its sign. The light shift of Rydberg levels have been measured, for example, for Xe atoms in an atomic beam \cite{OBrian1994} and for ultracold Rb atoms in an optical lattice \cite{Younge2010}. Theoretical calculations have been performed, for example, on the light shift for low-lying (n$<$9) states of rubidium \cite{Safronova2004} and approximate results for high lying Rydberg states can be found in Ref.\,\cite{Delone2000}. As the Rydberg states are close to the ionization threshold, the light field has usually enough energy to ionize the atom. Photoionization is therefore a possible interaction mechanism that will cause additional losses. The magnitude of both effects depends strongly on the wavelength and the intensity of the light field as well as on the principal quantum number $n$ of the Rydberg state.

Here, we report on the measurement of the AC-Stark shift of the 14$D_{5/2}$ Rydberg state of $^{87}$Rb in an optical dipole trap generated by a CO$_2$-laser. The atoms are initially prepared as a thermal cloud at mikrokelvin temperature. The experiment is performed in steady state by continuously exciting the atoms to the Rydberg state and looking at the production rate of rubidium ions upon photoionization by the CO$_2$-laser. We compare the observed spectra with a model that includes the AC-Stark shift and the finite lifetime against photoionization of the Rydberg state. For the chosen parameters both effects are very strong and easily visible in the experiment.

\section{Experimental setup}

The experiments are carried out on an apparatus for the production of ultracold quantum gases. Starting from a magneto-optical trap (MOT) we load $4\times 10^6$ rubidium atoms in a single beam optical dipole trap generated by a CO$_2$-laser with a waist of $30\,\mu$m. The initial power of the CO$_2$-laser is 10\,W, corresponding to a trap depth of $500\,\mu$K and a laser intensity of $7\times10^5$ W/cm$^2$ in the trap center. The atoms are prepared in the $|5S_{1/2},F=1\rangle$ hyperfine ground state and are equally distributed among all three Zeeman sublevels. We then ramp down the intensity of the CO$_2$-laser for evaporative cooling. After 6\,s we end up with a Bose-Einstein condensate of $10^5$ atoms at a final laser power of 50\,mW. In order to produce a thermal cloud we can stop the cooling ramp at any intensity in between. The temperature of the cloud is defined by the laser power and can be measured by standard absorption imaging. The spatial extension of the cloud is cigar-shaped and amounts to about $10\,\mu m \times 100\,\mu m$, slightly decreasing with temperature. After preparation, we keep the optical dipole trap at a constant power and switch on two additional light fields. The first light field is resonant with the $|5S_{1/2},F=2\rangle \leftrightarrow |5P_{3/2},F'=3\rangle$ transition of rubidium, which is also used for cooling and imaging the atoms. We refer to this laser as the ''imaging laser'' ($35\,\mu$W power, 5.4\,mm beam waist). As the atoms are initially in the $|5S_{1/2},F=1\rangle$ ground state, the imaging laser causes a weak off-resonant (6.8\,GHz detuning) optical pumping of atoms in the $|5S_{1/2},F=2\rangle$ ground state. The pumping rate is about 0.1\,s$^{-1}$. The second light field couples the $|5P_{3/2},F'=3\rangle$ excited state to the Rydberg state $|14D_{5/2}\rangle$ \cite{hyperfine}. It has a wavelength of 495\,nm and is generated via frequency doubling of a seeded diode laser in a periodically poled waveguide crystal. In the following, we refer to this laser as the ''Rydberg laser'' (1\,mW power, $250\,\mu$m beam waist). The extension of both lasers is larger than the extension of the atomic cloud and can be considered as homogeneous. The CO$_2$-laser, which provides the trapping potential, completes the three-photon ionization scheme. The relevant energy levels as well as the geometry of the laser beams are shown in Fig.\,1.

\begin{figure}[htbp]
\label{fig1}
\centering
\includegraphics[width=14cm]{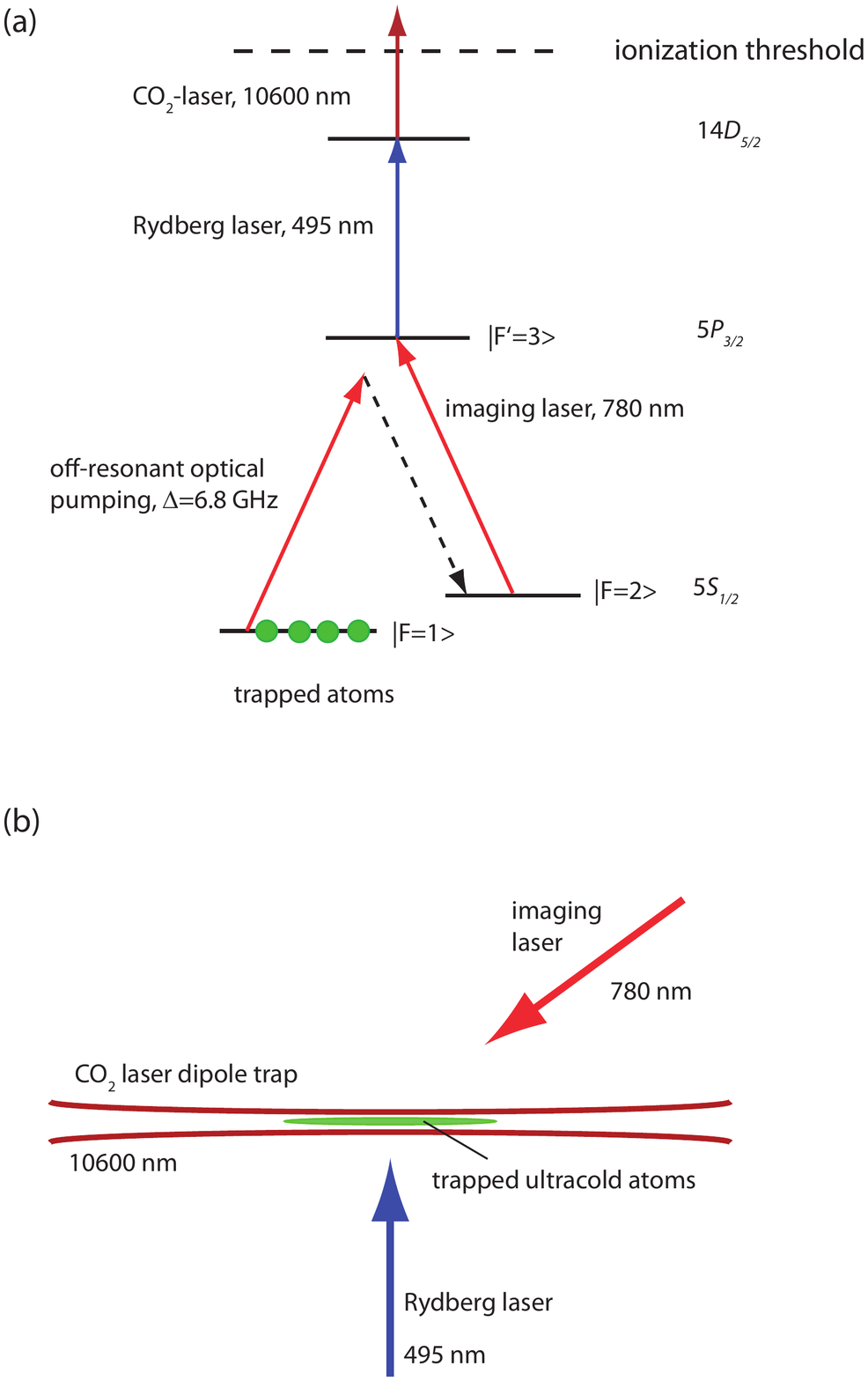}
\caption{Three-photon ionization scheme. (a) Relevant energy levels of rubidium. The imaging laser off-resonantly pumps the atoms from the $|5S_{1/2},F=1\rangle$ ground state to the $|5S_{1/2},F=2\rangle$ ground state. Subsequently, the atoms are ionized via the intermediate $|5P_{3/2},F'=3\rangle$ and the $|14D_{5/2}\rangle$ state. (b) Geometry of the three laser beams. The different directions are due to geometrical constraints of the vacuum system and have no special meaning. All laser beams are linearly polarized. The CO$_2$-laser provides both, the trapping potential and the final ionization step. The created ions are extracted with a small electric field (5\,$V/cm$) and counted by a channeltron detector.}
\end{figure}

{\bf Choice of the ${\bf 14D_{5/2}}$ state:} The experimental setup is part of a scanning electron microscope, which has been adopted for the imaging and manipulation of ultracold atoms. The detection principle relies on electron impact ionization of the atoms and subsequent ion detection \cite{Gericke2008,Wuertz2009}. The probability for electron impact ionization is less than 40\,\%. This currently limits the detection efficiency of this method. The dominant competing scattering channel is electron impact excitation. Thereby, most collisions lead to the excitation of the $5P_{1/2}$ and $5P_{3/2}$ states, as they have the largest dipole matrix elements. After excitation the atom eventually decays to the $|5S_{1/2},F=1\rangle$ or $|5S_{1/2},F=2\rangle$ ground state. As the atoms are initially in the $|5S_{1/2},F=1\rangle$ ground state, atoms that decay to the $|5S_{1/2},F=2\rangle$ ground state can be ionized by the above described three-photon-ionization scheme, thus enhancing the overall detection efficiency. The cross-section for photoionization of a Rydberg atom strongly depends on the binding energy \cite{Potvliege2006}. For the CO$_2$-laser wavelength, the {\it n}=13 state of rubidium is the lowest bound state that can be ionized. As the required wavelength of 498\,nm is inconvenient to generate we have decided to study photoionization via the $14D_{5/2}$ state at a wavelength of 495\,nm. For such low-lying Rydberg states one expects a very short lifetime against photoionization \cite{Potvliege2006}, resulting in a fast ionization scheme.

\section{Photoionization spectroscopy}

The signal that we use for the spectroscopy is the number of produced ions. We have recorded photoionization spectra for four different powers of the CO$_2$-laser. For each spectrum, the frequency of the Rydberg laser is varied while the frequencies of the imaging laser and the CO$_2$-laser are kept constant. For each setting of the Rydberg laser frequency we have performed one experimental run. We have chosen a total exposure time of 1\,s and have recorded the total number of detected ions during this time. Typically, a few thousand ions are detected, which is much less than the total number of atoms in the trap. Saturation effects can therefore be neglected. The total time of flight of an ion to the detector is 18\,$\mu$s (already after 1\,$\mu$s the ion has left the cloud) and the highest observed production rate of ions is 4000 s$^{-1}$. Therefore, there is almost never more than one ion or Rydberg atom at the same time inside the cloud and effects related to space charge, cold plasma formation, avalanche ionization or Rydberg blockade can be neglected. The spectra are shown in Fig.\,2. Each spectrum has two pronounced features. One rather sharp peak occurs close to resonance while a second peak with varying width is shifted with increasing intensity of the CO$_2$-laser. This second peak is due to the AC-Stark shift of the Rydberg state $14D_{5/2}$. 

\begin{figure}[htbp]
\label{fig2}
\centering
\includegraphics[width=10cm]{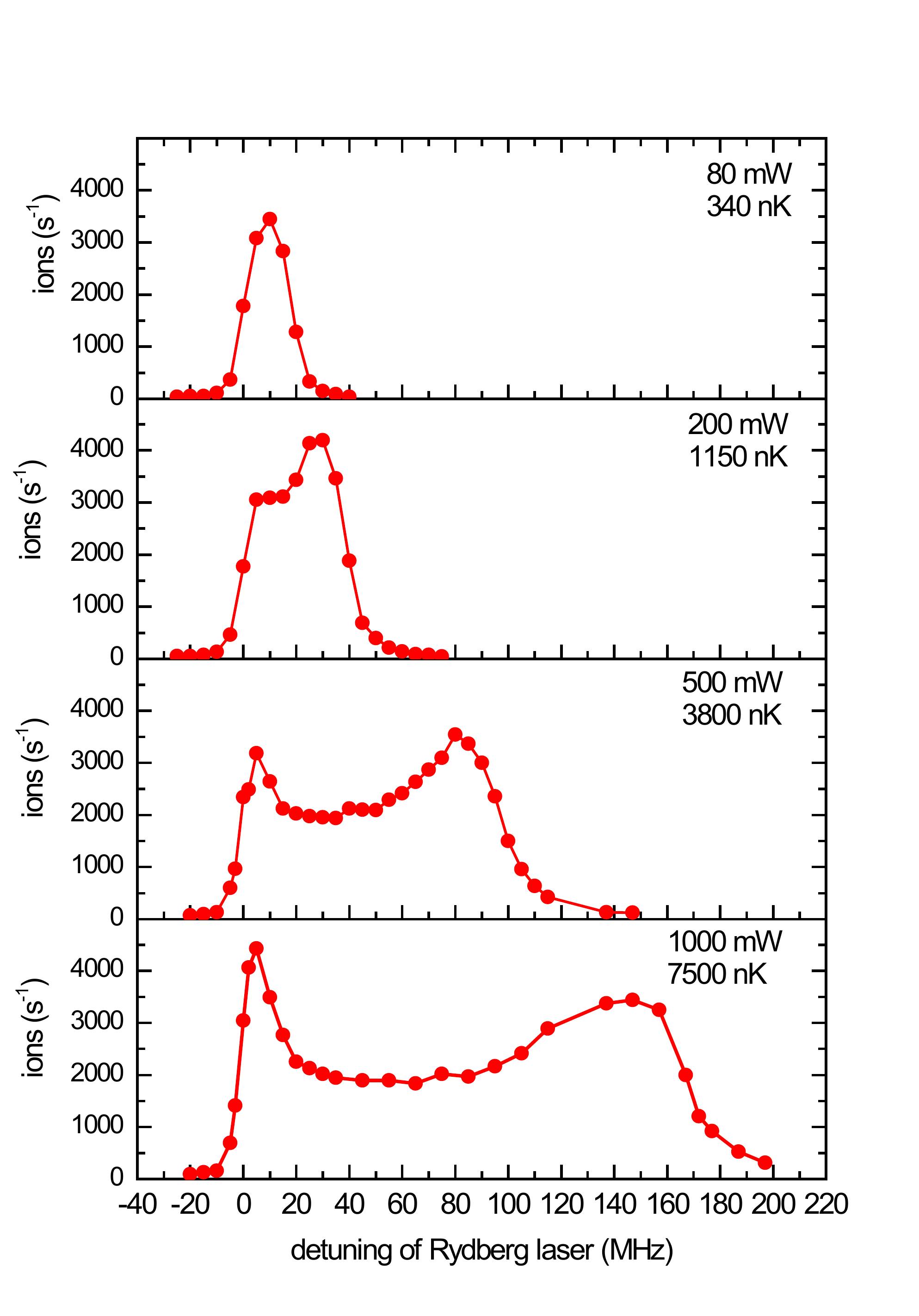}
\caption{Recorded photoionization spectra for different powers of the CO$_2$-laser. The laser power and the temperature of the thermal cloud are indicated in each graph. For zero detuning, the Rydberg laser is resonant with the $|5P_{3/2},F'=3\rangle \rightarrow |14D_{5/2}\rangle$ transition.}
\end{figure}

We start our analysis by first discussing the AC-Stark shifts, induced by the three involved lasers. The light-shift of the ground state $5S_{1/2}$ and the intermediate state $5P_{3/2}$ due to the CO$_2$-laser are well known \cite{Friebel1998}. The ground state is shifted towards negative energies as the laser is red detuned to all possible transitions. It is this shift that provides the trapping potential for the atoms. For the highest power of the CO$_2$-laser (1\,W) it amounts to $- h \times 1.1$\,MHz in the trap center. The shift of the $5P_{3/2}$ state is about two times as large and also negative. Compared to the size of the observed features in the spectra, both shifts can be neglected. We also note that the imaging laser is resonant for all four spectra as both light shifts are smaller than the natural linewidth of the intermediate $5P_{3/2}$.

The intensity of the imaging laser ($7 \times 10^{-5}$\,W/cm$^2$ in the beam center) ensures a small Rabi frequency of about $2\pi\times1$\,MHz. The Rydberg laser (1\,W/cm$^2$ in the beam center) couples the excited state to the Rydberg state with a Rabi frequency of $2\pi\times2.6$\,MHz. As the light shift cannot exceed the Rabi frequency, the imaging laser and the Rydberg laser do not induce considerable light shifts. The only remaining shift is that of the $14D_{5/2}$ state, induced by the CO$_2$-laser. 

\begin{figure}[htbp]
\label{fig3}
\centering
\includegraphics[width=10cm]{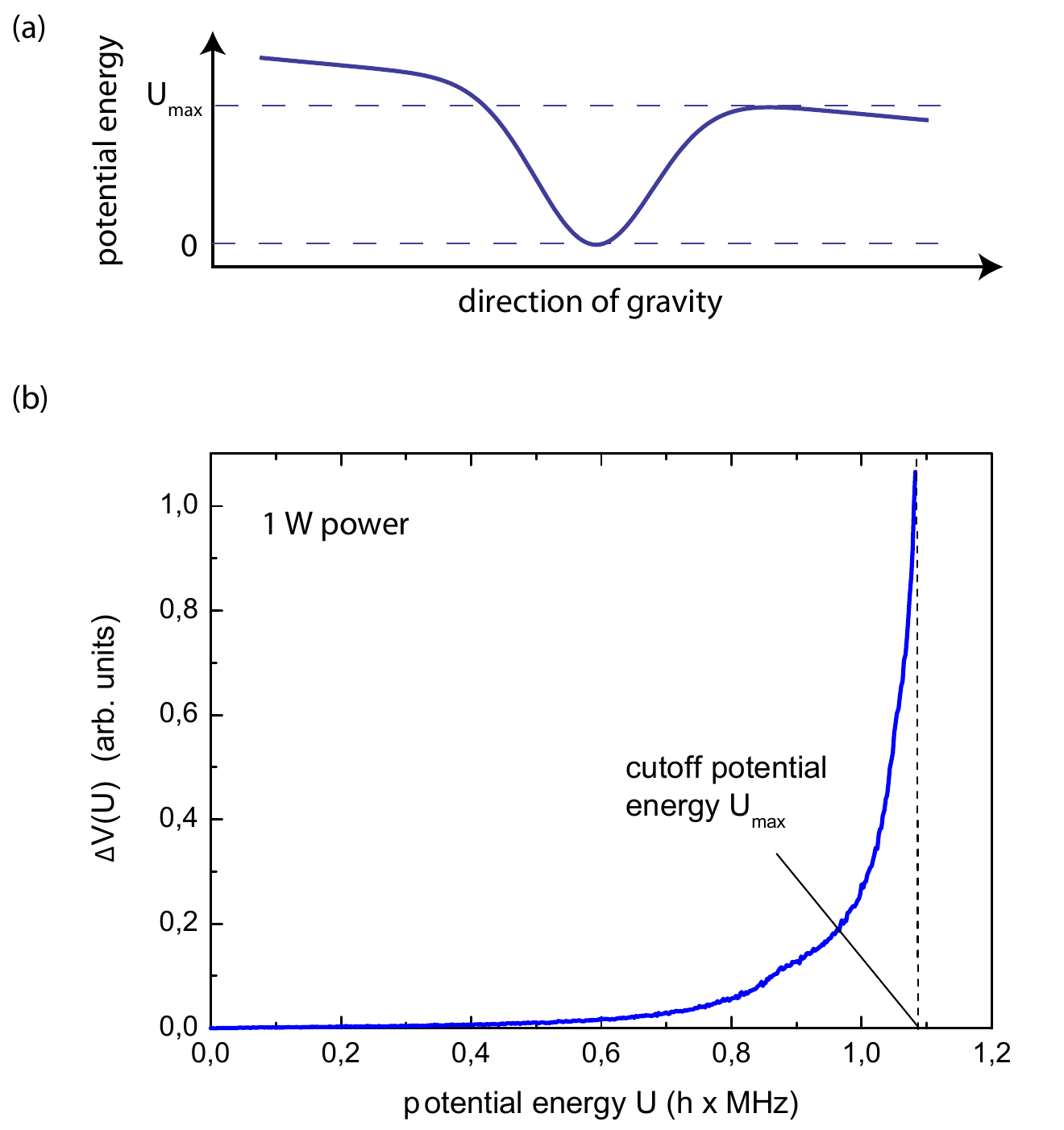}
\caption{(a) Trapping potential along the direction of gravity (not to scale). The saddlepoint defines the maximal potential energy $U_{\mathrm{max}}$ which we consider in our model. (b) Trap volume for a given potential energy, calculated for the spectrum with 1\,W power in the CO$_2$-laser. Due to the convex shape of the wings of the optical dipole trap, the volume becomes very large when the potential energy approaches $U_{\mathrm{max}}$.}
\end{figure}

The simplest way to model this shift is the ponderomotive potential of a free electron in the laser field. The potential energy corresponds to the average kinetic energy of an electron oscillating in the light field. It is given by

\begin{equation} \label{eq2}
h\nu_{\mathrm{ls}}=\frac{e^2I}{2m_e\epsilon_0 c(2\pi\nu_{\mathrm L})^2}\,\,,
\end{equation}

where $\nu_{\mathrm{ls}}$ denotes the resulting lightshift, $e$ is the electron charge, $I$ is the intensity of the light field at the position of the electron, $m_e$ is the electron mass, $c$ is the speed of light, $\epsilon_0$ is the dielectric constant, and $\nu_{\mathrm L}$ is the frequency of the laser. We have expressed the ponderomotive potential directly in terms of the lightshift. As the sign of the ponderomotive potential is always positive, the level is shifted upwards in energy and the light field constitutes a repulsive potential for the atom. Compared to a dipole trap with the same trap depth which is running at frequencies in the visible or near infrared spectrum, the ponderomotive potential in the CO$_2$-laser dipole trap is a factor of 1000 larger \cite{factor1000}.

\begin{figure}[htbp]
\label{fig4}
\centering
\includegraphics[width=10cm]{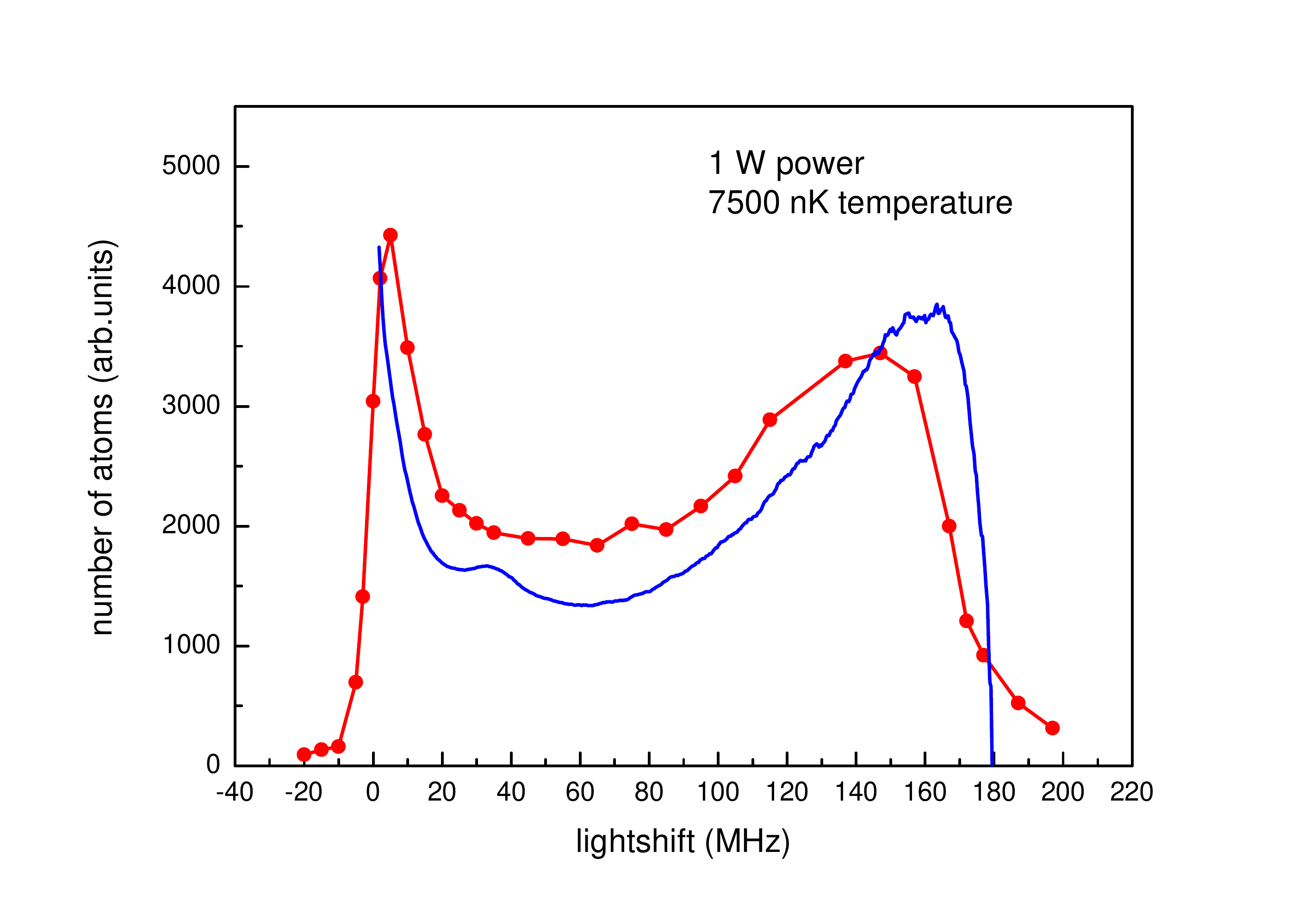}
\caption{Number of atoms with a given lightshift. The red points are the experimental data (see also Fig.\,2). The blue line is the calculated fraction of atoms with a given lightshift (normalized to the experimental data). The peak on the left corresponds to atoms close to the edge of the dipole trap, where the density is small but the available trap volume is large. The peak on the right stems from atoms in the trap center where the density is large but the trap volume is small. While there is already qualitative agreement, the shape of the peak at the right shows a significant deviation from the data.}
\end{figure}

Next, we have to take into account that in a dipole trap the lightshift of the Rydberg state depends on the position of the atom within the trap. For a thermal cloud, the density of the atoms is determined by the Boltzman distribution

\begin{equation}
n(\vec{r})\propto e^{-U(\vec{r})/k_{\mathrm B}T},
\end{equation}

where $U(\vec{r})$ is the trapping potential which is directly given by the intensity profile of the dipole trap laser: $U(\vec{r})\propto -I(\vec{r})$. Thus, according to Eq.\,\ref{eq2}, the lightshift of an atom is determined by its potential energy. The number of atoms with a certain lightshift is then identical to the number of atoms with a certain potential energy. This number is given by multiplying the density $n(\vec{r})$ with the available trap volume at this potential energy. We discretize the trap volume in potential energy shells of width $\Delta U$. The volume of each energy shell is then given by $\Delta V(U) =dV(U)/dU\times \Delta U$, where $V(U)$ is the integrated trap volume with a potential energy smaller than $U$. The calculation of $\Delta V(U)$ has to be done numerically. As an example, $\Delta V(U)$ for the spectrum with 1\,W power in the CO$_2$-laser is shown in Fig.\,\ref{fig3}b. Due to the asymptotic behaviour of the trapping potential, $\Delta V(U)$ diverges if the potential energy equals the trap depth. However, due to gravity, the symmetry is distorted and a saddle point of the potential along the direction of gravity emerges (see Fig.\,3a). We take the potential of the saddle point as a cutoff potential energy $U_{\mathrm {max}}$ for our calculation. 

In order to calculate the density one has to know the temperature of the cloud. In our approach, the temperature changes during the exposure as the evaporation continues after stopping the evaporation ramp. After 1\,s the temperature is about 30\,\% lower than at the beginning. We take the average of the initial and final temperature as the effective temperature for the measurement. The number of atoms with a certain lightshift is then readily calculated and plotted in Fig.\,4. It is clearly visible that a two-peak structure emerges, arising from the competition between the Boltzman distribution and the large number of available states at the edges of the dipole trap.

While the shape of the spectra is already visible, there is not yet full quantitative agreement. The reason is that the finite lifetime of the Rydberg state causes an additional broadening. The dominant contribution stems from photoionization. The lifetime of the Rydberg state depends on the photoionization cross-section $\sigma$ and the intensity of the CO$_2$-laser and is given by

\begin{equation}
\tau_{\mathrm{ion}}   =    \frac{h\nu_L}{I(\nu_{\mathrm {ls}})\sigma}\,\,.
\end{equation}

As the intensity is connected via Eq.\,1 to the lightshift $\nu_{\mathrm{ls}}$, the lifetime depends on the lightshift. The photoionization cross-section $\sigma$ of low-lying Rydberg states has been measured in Ref.\,\cite{Gabbanini2006}. For the $16D$ state it amounts to 39 Mb. Following the trend of the data we estimate a cross-section for the $14D_{5/2}$ state between 45 and 50 Mb. This corresponds to a lifetime in the trap center between 10 and 100 \,ns for the four different spectra. Assuming a Lorentzian profile with width $\delta \nu (\nu_{\mathrm{ls}})$ for the total broadening  we write  

\begin{equation}
\delta \nu (\nu_{\mathrm{ls}})= \delta \nu_{\mathrm{ion}}(\nu_{\mathrm{ls}}) + \delta \nu_{\mathrm{laser}} + \delta \nu_{\mathrm {natural}}\,\,,
\end{equation}

where $\delta \nu_{\mathrm{ion}} (\nu_{\mathrm{ls}})=(2\pi\tau_{\mathrm{ion}})^{-1}$ is the contribution from photoionization, $\delta \nu_{\mathrm{laser}}$ denotes a constant broadening due to the finite bandwidth of the lasers (1\,MHz each), and $\delta \nu_{\mathrm{natural}}$ denotes the natural linewidth of the Rydberg state (70\,kHz). 

\begin{figure}[htbp]
\label{fig5}
\centering
\includegraphics[width=10cm]{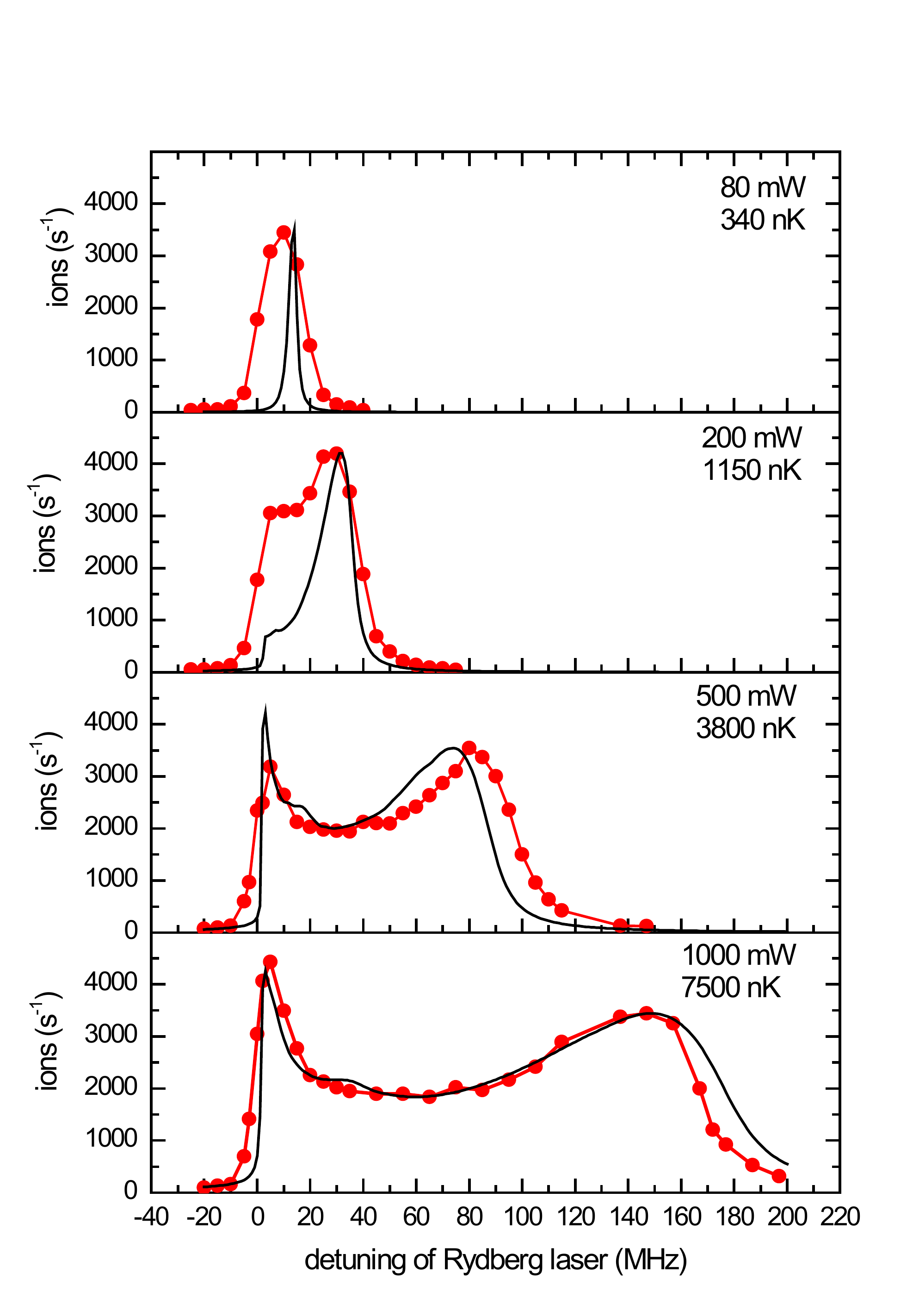}
\caption{Comparison with theory. The experimental data (red points, same data as in Fig.\,2) are shown together with the theoretical model as outlined in the text. The model has been normalized to the height of the shifted peak.}
\end{figure}

As all involved timescales (Rabi frequencies, lifetime and ionization rate) are much faster than the motion of the atoms in the trap, we further assume that the atoms are ionized right at the position where they are pumped into the $|5S_{1/2},F=2\rangle$ ground state. This allows to ignore the external dynamics of the atoms and to consider only a static density distribution. The final lineshape is then given by a convolution of the atom number distribution as shown in Fig.\,4 with the Lorentzian profile for the broadening.

In Fig.\,5 we show the result of the convolution together with the experimental data for a cross-section of 48\,Mb. The agreement is good, especially for a high power in the CO$_2$-laser. With the same parameters we can recover the shape of all spectra. Only the height and width of the left peak show stronger deviations. This is not surprising since the shape of the peak is very sensitive to the density at the edges of the dipole trap. As the evaporation is a dynamical process, there might be atoms in the trap that have a higher potential energy than $U_{\mathrm {max}}$ and therefore lead to a broadening of the unshifted peak. Moreover, the density at the trap edge is exponentially sensitive to the temperature which changes during the measurement. However, the position of the shifted peak is well described for all data sets. This is important as this peak contains the information about the lightshift and the photoionization cross section. 

Spontaneous decay (2.2\,$\mu$s lifetime) and transitions induced by black body radiation (10\,$\mu$s lifetime) can cause a redistribution of the $14D_{5/2}$ state to neighboring states. As the ionization process takes place on a timescale which is at least 20 times faster, it is sufficient to restrict the analysis to the $14D_{5/2}$ state. Also, ionization due to black body radiation does not play a significant role as it amounts to only a fraction of the rate for black body induced transitions \cite{Glukhov2010}. Electric fields are another possible source of line broadening and line shifts. Our measurement principle requires a small electric field (5\,V/cm) which is continuously applied during the experiment. While such a field can significantly shift high-lying Rydberg states, its influence on the $14D_{5/2}$ state is less than 1\,MHz, which is below the resolution of our spectroscopy technique.

We conclude the discussion by a detailed analysis of the validity of the ponderomotive potential. The assumption of a free electron for the $14D_{5/2}$ state is certainly questionable as the binding energy corresponds to 70\,\% of the photon energy and resonance effects might occur. A quantum-mechanical calculation is therefore necessary for a verification. This is most conveniently done by writing the interaction of the electron with the radiation field in terms of the vector potential \cite{vectorpotential}

\begin{equation}
H_\mathrm{int}(t)=\frac{e^2}{2m}{\bf A}(t)^2+\frac{e}{m}{\bf A}(t){\bf p},
\end{equation}

with ${\bf p}$ being the electron momentum and ${\bf A}(t)=-{\bf E_0}/\omega_{\mathrm L}\cos(\omega_{\mathrm L} t)$, where ${\bf E_0}$ is the electric field vector of the light field. The first term directly gives the ponderomotive potential in first order perturbation theory after time averaging over one oscillation period, $\Delta E_1=e^2E_0^2/(4m\omega_{\mathrm L}^2)$. It shifts all states in the same way. The second term can then be regarded as a correction to the ponderomotive potential. In second order perturbation theory one can write

\begin{equation}
\label{correction}
\Delta E_2 =\frac{e^2E_0^2}{4m\omega_{\mathrm L}^2}\times\frac{1}{\hbar}\sum_k\frac{|\left\langle k|z|i\right\rangle|^22m\omega_{ik}^3}{\omega_{ik}^2-\omega_{\mathrm L}^2}.
\end{equation}

Here, we have set the linear polarization of the light field along the $z$-axis and have replaced the matrix elements according to $\left\langle k|{\bf p}|i\right\rangle=im\omega_{ik}\left\langle k|{\bf r}|i\right\rangle$, with $\hbar\omega_{ik}=E_i-E_k$, $E_i$ and $E_k$ being the energies of the initial state $|i\rangle$ and the intermediate states $|k\rangle$. The first factor in Eq.\,\ref{correction} is again the ponderomotive potential and the second factor is a dimensionless correction factor. The $14D_{5/2}$ state is coupled to all $nP_{3/2}$, $nF_{5/2}$ and $nF_{7/2}$ states and we have included in the calculation all intermediate states from $n=5$ to $n=120$. Note that 90 percent of the lightshift originates from the states up to $n=40$. The wave functions have been generated with help of the Numerov method and the quantum defects have been taken from Ref.\,\cite{Lorenzen1983}. The calculation has been performed for $|m|=1/2, 3/2$, and $5/2$, where $m$ is the projection of the total angular momentum on the electric field vector of the CO$_2$-laser. For all three Zeeman sub-states the correction factor $c_m$ to the pondermotive potential is less than 10 percent. We find $c_{1/2}=0.07$, $c_{3/2}=0.05$, and $c_{5/2}=0.02$. In the experiment we populate a mixture of all three sublevels. In Fig.\,6 we show the level shift arising from the coupling to the $nP_{3/2}$ states for $|m|=1/2$. It is clearly visible that a peak-like structure appears around $n=11$, where the CO$_2$-laser is close to resonance. However, the detuning is still large enough to ensure a ponderdomotive potential. Note that the contributions from the various states partially cancel. 

\begin{figure}[htbp]
\label{fig7}
\centering
\includegraphics[width=10cm]{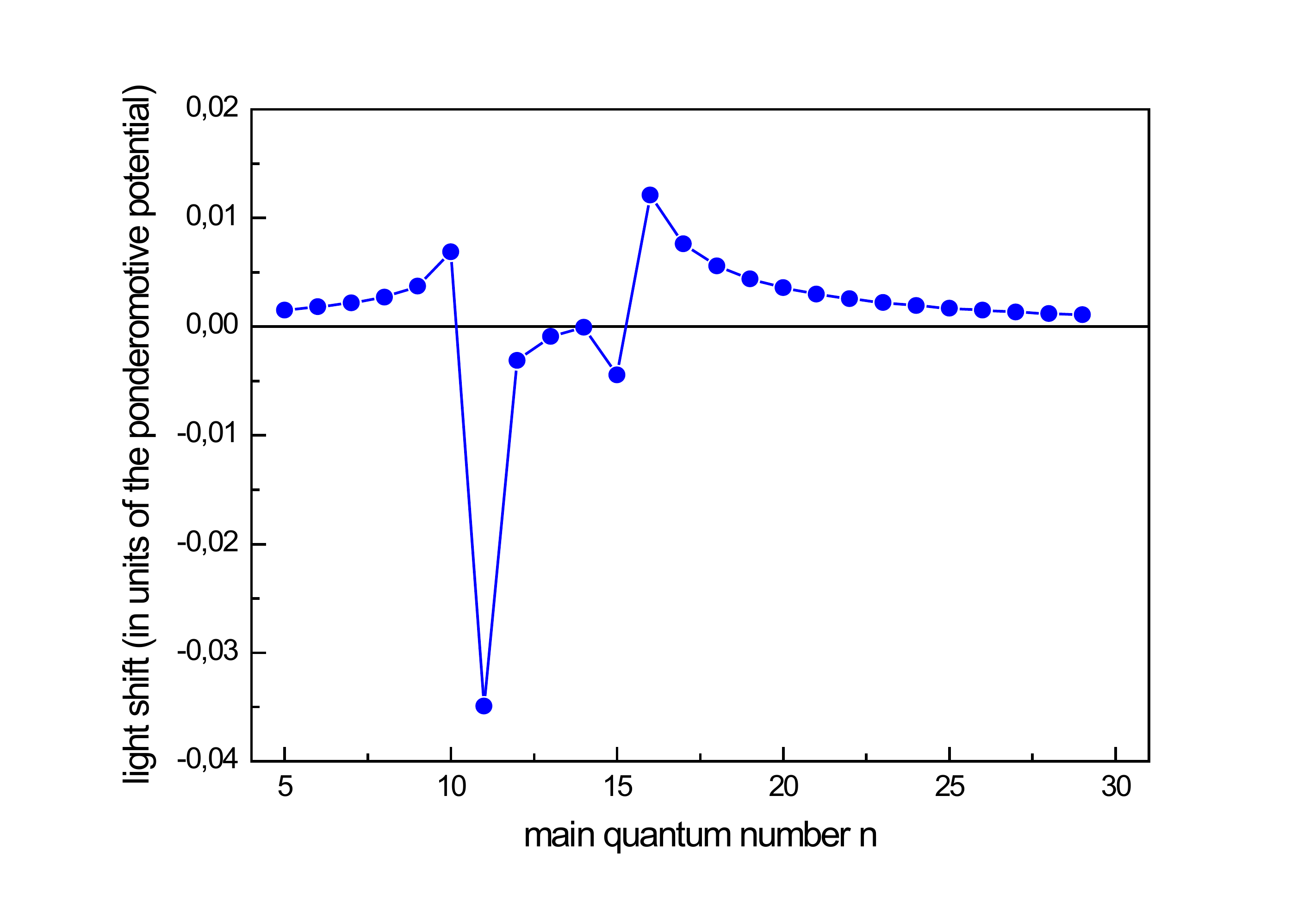}
\caption{Contribution to the light shift from the intermediate $nP_{3/2}$ states for $|m|=1/2$, see Eq.\,\ref{correction}. The light shift is given in units of the ponderomotive potential.}
\end{figure}

\begin{figure}[htbp]
\label{fig6}
\centering
\includegraphics[width=9cm]{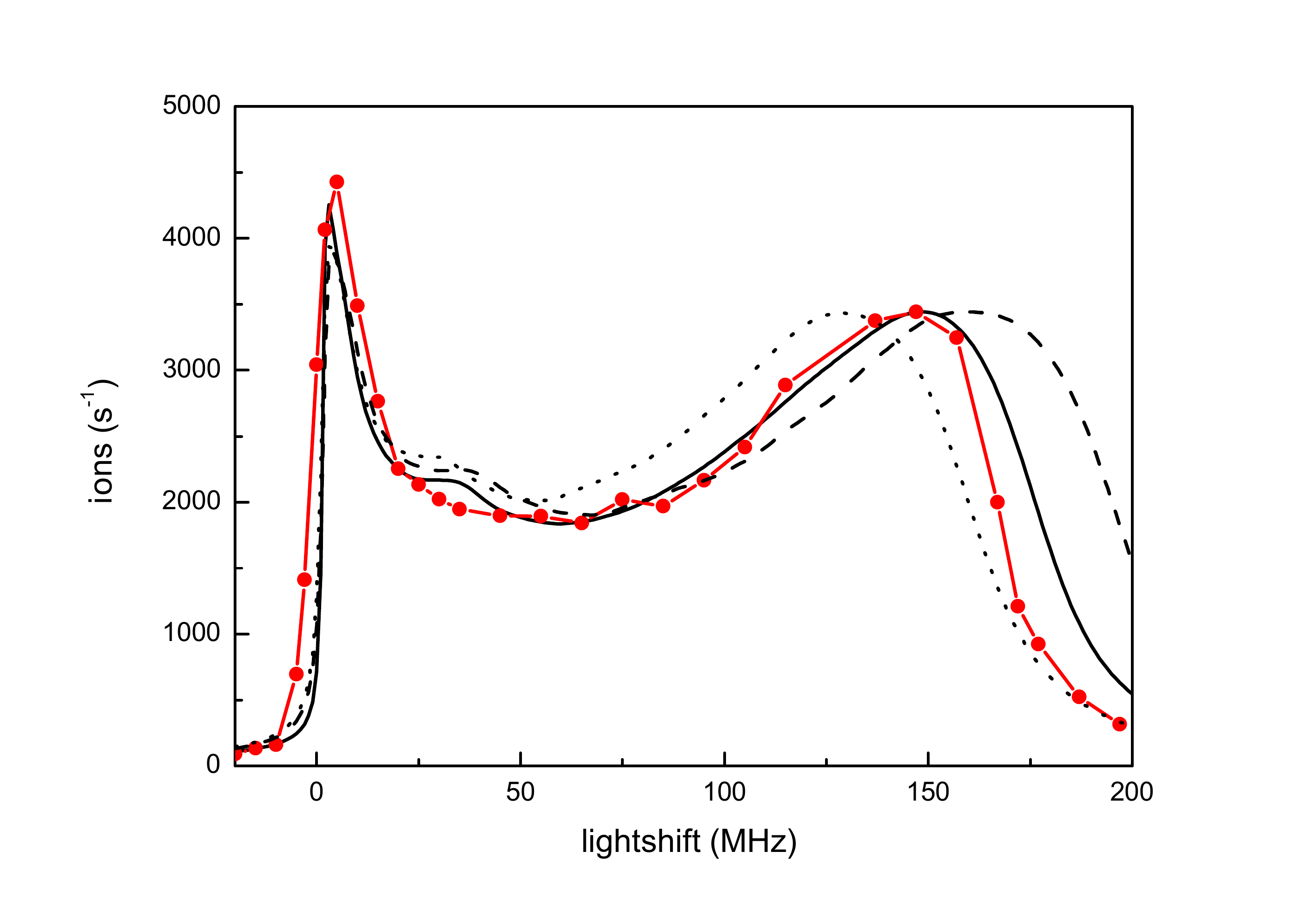}
\caption{Comparison of the 1\,W spectrum with different strengths of the lightshift. The dotted (dashed) line corresponds to a lightshift with 10\,\% less (10\,\% more strength.)  }
\end{figure}

The above presented model has (apart from the normalization constant) no free parameter. In order to test for a possible deviation from the ponderomotive potential we can artificially tune the strength of the ponderomotive potential with an additional factor $\eta$ and repeat the evaluation for different values of $\eta$. This is shown in Fig.\,7 for $\eta$=0.9 and 1.1. As one can see, the deviation of 10\,\% already leads to a disagreement with the observed spectra. This is in accordance with the detailed calculation and we can conclude that the AC-Stark shift of the $14D_{5/2}$ state in a CO$_2$-laser dipole trap is given by the ponderomotive potential of a free electron. A similar result has been obtained for low-lying Rydberg states of Xenon (n=10,...,15) which were also found to be in good agreement with a ponderomotive potential \cite{OBrian1994}.

\section{Summary and Outlook}

We have measured the AC-Stark shift of the $14D_{5/2}$ state of rubidium in a CO$_2$-laser dipole trap. We find that the lightshift is given by the ponderomotive potential of a free electron in the light field. The ponderomotive potential is always repulsive and is independent of the principal quantum number $n$. All higher lying Rydberg states are shifted in the same way, provided that no near-resonant coupling to lower lying states occurs. For our settings we observe a light shift of up to 170\,MHz. This can be used, for instance, for new schemes of evaporative cooling, as the excitation of the atoms to the Rydberg state can be made spatially selective. We also extract the photoionization cross-section from our data which we find to be compatible with previous measurements. The observed short lifetime of the Rydberg state of less than 100\,ns even for a shallow trapping potential sets a limitation for the use of low-lying Rydberg states in combination with a CO$_2$-laser dipole trap. However, for higher quantum numbers, the ionization cross-section drastically decreases and lifetimes in the ms range are realistic \cite{Potvliege2006}. Both effects, the light shift and the lifetime against photoionization can be significantly reduced using dipole traps in the visible or near-infrared spectral range. This will make experiments with Rydberg-dressed atoms in optical dipole traps feasible.


\begin{thebibliography}{10}
\bibitem{hyperfine}
The hyperfine structure of the Rydberg state is not resolved in our experiment and is therefore neglected.
\bibitem{factor1000}
The $1/\nu^{2}$-dependence of the ponderomotive potential leads to a factor of 100. As the trapping potential for a ground state atom is proportional to $I/\Delta$, where $\Delta$ is the detuning from resonance an additional factor of 10 occurs as the intensity has to be increased by this factor in order to get the same potential.
\end{thebibliography}
\end{document}